\documentclass[a4paper,10pt]{article}

\usepackage[english]{babel}
\usepackage[latin1]{inputenc}
\usepackage{amsmath,amssymb,amsthm,amsfonts}
\usepackage{graphicx}
\usepackage{enumerate}
\usepackage{hyperref}
\usepackage{a4wide}
\usepackage{stmaryrd}

\newcommand{\comment}[1]{}	
\newcommand{\R}{\mathbb{R}}

\newcommand{\D}{\Delta}
\newcommand{\e}{\varepsilon}
\renewcommand{\a}{\alpha}
\renewcommand{\d}{\delta}
\renewcommand{\epsilon}{\varepsilon}

\newcommand{\mA}{\vert A \vert}
\newcommand{\mB}{\vert B \vert}
\newcommand{\md}{\mu_\d}
\newcommand{\mt}[1][t]{\vert #1(A) \cap B \vert}
\newcommand{\dBall}[1][t]{B_\d^{#1}}
\newcommand{\prob}[1][t]{P(\dBall[#1])}
\newcommand{\topt}{t^{\mathrm{opt}}}
\newcommand{\tdpt}{t^{\mathrm{app}}}
\newcommand{\tapp}{t^\ast}

\newtheorem{theorem}{Theorem}

\newtheorem{lemma}[theorem]{Lemma}
\newtheorem{cor}[theorem]{Corollary}
\newtheorem{prop}[theorem]{Proposition}

\title{Probabilistic Matching of Planar Regions\footnote{
This work was partially supported by the European Union under contract No. FP6-511572, Project PROFI and  by the DFG Priority Programme 1307 \emph{Algorithm Engineering}. 
}}
\author{Helmut Alt \and Ludmila Scharf \and Daria Schymura}
\date{Institut f\"ur Informatik, Freie Universit\"at Berlin}

\begin{document}

\bibliographystyle{plain}
\maketitle

\begin{abstract}
We analyze a probabilistic algorithm for matching shapes modeled by planar regions 
under translations and rigid motions (rotation and translation).
Given shapes $A$ and $B$, the algorithm computes a transformation $t$
such that with high probability the area of overlap of $t(A)$ and $B$
is close to maximal.
In the case of polygons, we give a time bound that does not depend
significantly on the number of vertices.
\end{abstract}

\section{Introduction}

\paragraph{The Problem.}

Matching two geometric shapes under transformations and evaluating their similarity is one of the central problems in computer vision systems where the evaluation of the resemblance of two images is based on their geometric shape and not color or texture. Because of its significance the problem has been widely covered in the literature, 
see~\cite{alt99discrete,Veltkamp:fk} for surveys.

Depending on the application, 2D shapes are modeled as finite point patterns, polygonal chains or polygons.
Given two shapes $A$ and $B$, as well as a set of transformations $T$ and a distance measure $d$, the problem is to find the transformation $t \in T$ such that $t(A)$ and $B$ match optimally with respect to $d$.
Two shapes are considered similar if there is a transformation $t$ such that the distance between $t(A)$ and $B$ is small.
The problem is well-studied for various settings, e.g., sets of line segments, rigid motions and the Hausdorff distance. 

In this paper we consider the problem of matching 2D shapes modeled by plane open sets, e.g., sets of polygons, with respect to the area of the symmetric difference, which is the area that belongs to exactly one of the shapes.
As  sets of allowed transformations $T$ we will consider the set of translations and the set of rigid motions (rotation and translation) in the plane.
Minimizing the area of the symmetric difference under translations or rigid motions is equivalent to maximizing the area of overlap, so we will consider the latter formulation of the problem for the rest of this article.
The area of overlap is a well-known similarity measure, and, e.g., has the advantage that it is insensitive to noise.
Furthermore, computing the maximal area of overlap of two sets of polygons under translations or rigid motions is an interesting computational problem on its own.

\paragraph{Related Work.}

For simple polygons, efficient algorithms for maximizing the area of overlap
under translations are known.
Mount et al.~\cite{mount} show that the maximal area of overlap of a simple $n$-polygon with a translated simple $m$-polygon can be computed in $O(n^2 m^2)$ time.
Recently, Cheong et al.~\cite{finding_a_guard} introduced a general probabilistic framework for computing an approximation with prespecified absolute error $\e$ in $O(m+(n^2/\e^4)\log(n)^2)$ time for translations and $O(m+(n^3/\e^4)\log(n)^5)$ time for rigid motions.

De Berg et al.~\cite{deBerg} consider the case of convex polygons and give a $O((n+m)\log(n+m))$ time algorithm  maximizing the area of overlap under translations.
Alt et al.~\cite{constant_factor} give a linear time constant factor approximation algorithm for minimizing the area of the symmetric difference of convex shapes under translations and homotheties (scaling and translation).

For higher dimensions
Ahn et al.\ present in~\cite{Ahn:2008jx} an algorithm  finding a translation vector maximizing the overlap of two convex polytopes bounded by a total of $n$ hyperplanes in $\R^d$ for $d\ge 3$. Their algorithm runs in $O(n^{\lceil d/2 \rceil+1}\log^{d-1} n)$ time with probability at least $1-n^{-O(1)}$.

Surprisingly little has been known so far about maximizing the area of overlap under rigid motions.

\paragraph{Overview.}
We will design and analyze a simple probabilistic matching algorithm, which for translations works as follows.
Given two shapes $A$ and $B$, in one random experiment we select a point $a\in A$ and a point $b \in B$ uniformly at random. This tells us that the translation $t$ that is given by the vector $b-a$ maps some part of $A$ onto some part of $B$. We record this as a vote for $t$ and repeat this procedure very often. Then we  determine the densest cluster of the resulting point cloud and output the center of this cluster as a translation that maps a large part of $A$ onto $B$. For rigid motions we consider two different approaches for the vote generation in one random experiment.

We show that the  algorithm  approximates the maximal area of overlap under translations and rigid motions. More precisely, 
let $\topt$ be a transformation that maximizes the area of overlap of $A$ and $B$, and let $\tapp$ be a transformation computed by the algorithm.  Given an allowable error $\e$ and a desired probability of success $p$, both between $0$ and $1$, we show bounds on the required number of random experiments, guaranteeing that the absolute difference between approximation and optimum $\mt[\topt] - \mt[\tapp]$ is at most $\epsilon \mA$ with probability at least $p$.
Here $\vert \cdot \vert$ denotes the area (Lebesgue measure) of a set.
Furthermore, we prove that this algorithm computes a $(1+\e)$-approximation of the maximal area of overlap under translations and rigid motions, meaning that $\mt[\topt] - \mt[\tapp] \leq (1+\e) \mt[\topt]$ with high probability, if we make a reasonable assumption about the input shapes.

This algorithm is a special case of a probabilistic algorithmic scheme for approximating an optimal match of planar sets under a subgroup of affine transformations.
Alt and Scharf \cite{09-smrs} analyzed another instance of this algorithmic scheme that compares polygonal curves under translations, rigid motions, and similarities.

\section{The Algorithms}\label{sec:algorithms}

\subsection{Description of the Algorithms}
\label{sec:alg}
\paragraph{Shapes.}
We consider  shapes modeled by open bounded, and therefore,  Lebesgue measurable subsets of the plane. 
We always assume the shapes to have positive area. 
Additionally, we assume that there is a method to select uniformly distributed random points from a shape and the density function is Lipschitz continuous (see Section \ref{sec:lipschitz}).
This is the case for sets of  disks and for sets of polygons, or equivalently, sets of triangles, which probably is the most common representation in practice.
The idea of the algorithm can be  applied to bitmap data as well.

For a shape represented by $n$ triangles a random point can be generated by first selecting a  triangle randomly with probability proportional to the relative area of the triangle and then selecting a random point from that triangle.

For an arbitrary Lebesgue measurable set $A$ in the plane if we are given a set of triangles $\mathcal{T}$ it is contained in, and an algorithm that decides ``Is $p \in A$?'', we can sample uniformly at random from $\bigcup \mathcal{T}$ and discard points that are not in $A$.
The density function is Lipschitz continuous, for example, if the shape boundaries are unions of piecewise differentiable simple closed curves. 

\paragraph{General Idea.}
The idea of the algorithm is quite simple. Given two shapes $A$ and $B$, repeat the following random experiment very often, say $N$ times: Select random point samples of appropriate size from each shape and compute a transformation that maps the point sample of one shape to the  sample  of the other shape. Keep this transformation, called a ``vote'', in mind.
In each step, we grow our collection of ``votes'' by one. Clusters of ``votes'' indicate transformations that map large parts of the shapes onto each other.

Every translation can be associated with a point in two dimensional space  and every rigid motion with a point in three dimensional space. The densest cluster of ``votes'' is then defined as the transformation $t^* $ whose $\delta$-neighborhood with respect to the maximum norm contains the most transformation points from random experiments for some parameter $\delta$\ \footnote{In order to make it reasonable to use the same tolerance $\delta$ for both, translations and rotations, it is advisable to normalize the translation space.}.
Thus, along with the shapes $A$ and $B$ we have two additional input parameters: $N$ determines the number of random experiments and $\d$ adjusts the clustering size.

\begin{figure}[h]
\begin{center}
\includegraphics[width=0.6\columnwidth]{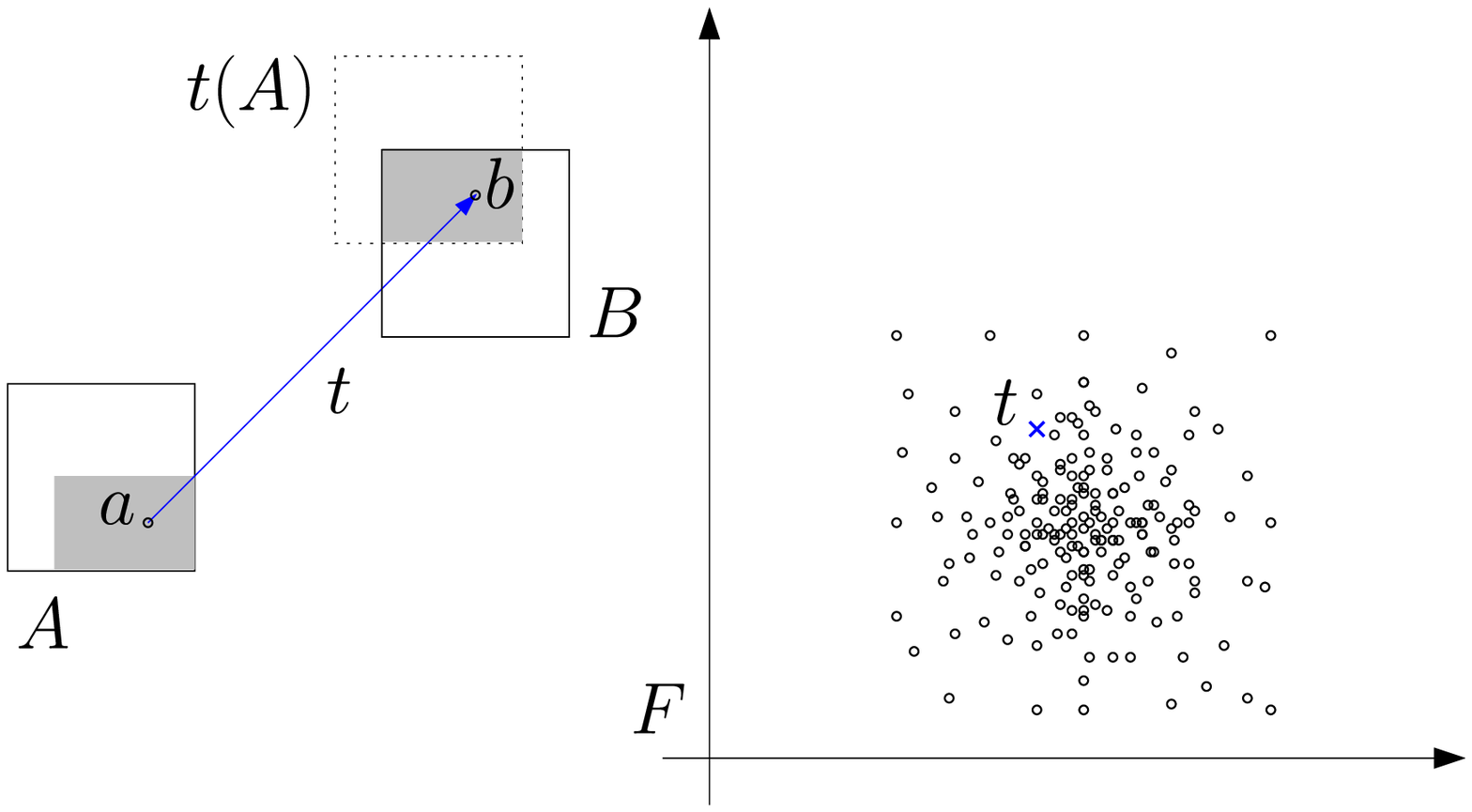}
\caption{We compare two copies of a square under translations.
The area of overlap of $t(A)$ and $B$ corresponds to the chance
of choosing a point pair $(x,y) \in A \times B$ such that $y-x=t$.}
\label{fig:two_squares}
\end{center}
\end{figure}

This  algorithm captures the intuitive notion of matching. Transformations whose $\delta$-neigh\-bor\-hoods contain many ``votes'' should be ``good'' translations since they map many points from $A$  onto points from $B$.
Figure \ref{fig:two_squares} illustrates this idea for the case of translations.

\paragraph{Translations.}
Observe that two points in the plane uniquely determine  a translation that maps one point onto the other. Therefore, a point sample for the case of translations consists of one randomly selected point of each shape.

\smallskip \noindent
\textsc{ProbMatchT}\\ 
\textsc{Input: }shapes $A$ and $B$, an integer $N$, and a positive real $\d$.	
\begin{enumerate}
   \item Perform the following experiment $N$ times:\\ 
      Draw uniformly distributed random points $a \in A$ and $b \in B$.\\ 
      Register the translation vector $b-a$.
   \item Determine a translation $\tapp$ whose $\d$-neighborhood contains the most registered vectors.
\end{enumerate}
\textsc{Output: } translation $t^\ast$
\smallskip

\paragraph{Rigid Motions with Random Angle.}

The algorithm for rigid motions, \textsc{ProbMatchRMRA},
is similar to the algorithm for translations.
The space of rigid motions $R$ is given as 
$I \times \nolinebreak \R^2 \subset \R^3$ where $I =[-1/2,1/2)$. 
We use the interval $[-1/2,1/2)$ instead of $[-\pi,\pi)$
because we regard this interval as a probability space,
which should have measure 1, avoiding a constant in the density function.
A point $(\alpha,t) \in R$ denotes the rigid motion
\[
x \mapsto M_{\alpha} x + t, \quad M_\alpha = \begin{pmatrix}
\cos 2\pi \a & - \sin 2\pi \a   \\ \sin 2\pi \a   & \cos 2\pi \a \end{pmatrix}.
\]

For matching under rigid motions, we select in each step uniformly
distributed an angle $\a$ and random points $a \in A$ and $b \in B$.
We give one ``vote'' to the unique rigid motion
with counterclockwise rotation angle $\a$ that maps $a$ onto $b$, namely the map
\[
x \mapsto M_{\alpha} x + (b - M_{\alpha} a).
\]

\paragraph{Rigid Motions with 3+1 Points.}

Another  variant for rigid motions is the algorithm \textsc{ProbMatchRM3+1}, which does not choose a completely random rotation but prefers directions that are present in the shape.
A rigid motion is determined by selecting two points $a_{1},a_{2}$ in $A$ and one point $b_{1}$ in $B$ uniformly at random.
Then, we select another point $b_{2}$ in $\R^2$ such that the distances between the points in $a_{1}$ and $a_{2}$ and $b_{1}$ and $b_{2}$ are the same, i.e., $b_2=b_{1}+\Vert a_{2}-a_{1}\Vert M_{\beta}\genfrac{(}{)}{0pt}{1}{1}{0}$, where $\beta\in[-1/2,1/2)$ is randomly selected under uniform distribution.
If $b_{2}$ happens to be in $B$, $(a_{1},a_{2},b_{1},b_{2})$ is a valid random sample. Otherwise,
we discard the sample and select new points.
In this way, we select uniformly distributed tuples from
\[
\mathcal{S}=\{ (a_{1},a_{2},b_{1},\beta) \in A^2 \times B \times I :
b_{2}=b_{1}+ \Vert a_{2}-a_{1}\Vert M_{\beta} \genfrac{(}{)}{0pt}{1}{1}{0} \in B \}.
\]

\subsection{Main Results}
\label{sec:mainRes}
\paragraph{Approximation Theorems.}
First, we give bounds on the required number of random experiments.
The main results are the following approximation theorems.

\begin{theorem}[Absolute Approximation]\label{thm:main_absolute}
For any two shapes $A$ and $B$ and  parameters $\e,\tau$ with $0<\e,\tau<1$ there exist a positive real $\d=\Theta(\e)$ and an integer $N$ such that the following holds: 
Let the transformation $t^\ast$ be the output of \textsc{ProbMatchT} or \textsc{ProbMatchRMRA}, respectively, let $\topt$ be a transformation that maximizes the area of overlap of $A$ and $B$, then 
\[
\left| \ \mt[t^\ast] - \mt[\topt] \ \right| < \e \mA
\]
with  probability at least $1-\tau$. 
In the case of translations
\[
N=O\left( c / \e^6 \log ( \max\left\{ 1/\tau,\ c / \e \right\} ) \right)\enspace ,
\]
in the case of rigid motions
\[
N=O\left( C / \e^8 \log  ( \max\left\{ 1/\tau,\ C / \e \right\} ) \right)\enspace ,
\]
where  $c=\mB^2 \Delta^4/\mA^4 $, $C=\mB^2 \Delta^6 D^6 / \mA^6$,
$\Delta$ is the length of the boundary of $A$  and $D$ is the diameter of $A$.
\end{theorem}

If we know that the shapes we have to match are not too ``skinny'' we can also bound the number of experiments required by algorithm \textsc{ProbMatchRM3+1} in order to achieve the absolute approximation error of at most $\e \mA$.

We say that a shape $A$ is \emph{$\kappa$-fat}\ \footnote{Our definition of $\kappa$-fatness differs from the standard definition.} for some constant $0<\kappa\le 1$ if there exists an inscribed circle $\mathcal{C}\subset A$ such that $|\mathcal{C}|\ge\kappa\mA$.

\begin{theorem}
   Let $A$ and $B$ be two $\kappa$-fat shapes such that the largest inscribed circle in $A$ is at most as large as the largest inscribed circle in $B$. For all parameters $\e,\tau$ with $0<\e,\tau<1$ there exist a positive real $\d=\Theta(\e)$ and an integer $N$ such that the following holds: 
Let the transformation $t^\ast$ be the output of \textsc{ProbMatchRM3+1} and $\topt$  a transformation that maximizes the area of overlap of $A$ and $B$, then 
\[
\left| \ \mt[t^\ast] - \mt[\topt] \ \right| < \e \mA
\]
with  probability at least $1-\tau$. 
The required number of experiments $N$ is
\[
N= O\left(  \max\left\{  \frac{C'}{\e^8 \kappa^5}  \log \left( \max\left\{ \frac{1}{\tau},\ \frac{C'}{\e \kappa} \right\} \right), \ \frac{1}{\kappa^6} \log \frac{1}{\tau} \right\} \right)
\]
where   $C'= \mB^2 \Delta^6 D^6 / \mA^{6} $,
$\Delta$ is the length of the boundary of $A$ and $D$ is the diameter of $A$.
   \label{thm:abs31}
\end{theorem}

Further, under the assumption that the shapes are $\kappa$-fat, we get a relative approximation for all three variants of the algorithm from the absolute approximation results.
For  algorithms \textsc{ProbMatchT} and \textsc{ProbMatchRMRA} a weaker assumption that the maximal area of overlap of $A$ and $B$ is at least a constant fraction of $\mA$, i.e., $\max_t \mt \geq \kappa \mA$ for some $0<\kappa\le 1$,  is sufficient for the relative error bound. Observe that if the two shapes $A$ and $B$ are $\kappa$-fat and $A$ is the shape with the smaller largest inscribed circle, then $\max_t \mt \geq \kappa \mA$.

The relative error bound follows if we choose $\e' = \e  \kappa$ and apply the absolute approximation results:
\[
\vert \mt[\topt] - \mt[\tapp] \vert \le \e' \mA = \e\kappa\mA  \le \e \mt[\topt].
\]

\begin{cor}[Relative Approximation]
   Given two shapes $A$ and $B$, let  $t^\ast$, $\topt$, $\e$, $\delta$, $\tau$, $N$ be as in Theorem~\ref{thm:main_absolute} for   algorithms \textsc{ProbMatchT} and \textsc{ProbMatchRMRA}, and as in Theorem~\ref{thm:abs31} for algorithm \textsc{ProbMatchRM3+1}.
Assume that $\mt[\topt] \geq \kappa \mA$ for some constant $\kappa$ in case of translations and rigid motions with random rotation angle. For rigid motions with $3+1$ points assume that $A$ and $B$ are $\kappa$-fat.
Then with probability at least $1-\tau$
\[
\left| \mt[t^\ast] - \mt[\topt] \right| < \e \mt[\topt]
\]
if $N$ is chosen as in Theorem~\ref{thm:main_absolute} for  algorithms \textsc{ProbMatchT} and \textsc{ProbMatchRMRA}, only that now   $c=\mB^2 \Delta^4 / \mA^4 \kappa^6$ and $C= \mB^2 \Delta^6 D^6/ \mA^6 \kappa^8$, where $\Delta$ is the length of the boundary of $A$ and $D$ is the diameter of $A$.
For algorithm \textsc{ProbMatchRM3+1} the necessary number of experiments is
$N= O\left(  \max\left\{  \frac{C'}{\e^8 \kappa^{13}}  \log \left( \max\left\{ \frac{1}{\tau},\ \frac{C'}{\e \kappa} \right\} \right), \ \frac{1}{\kappa^6} \log \frac{1}{\tau} \right\} \right)$, where $C'$ is as in Theorem~\ref{thm:abs31}.
\end{cor}

\paragraph{Runtime for Sets of Polygons.}
The runtime of the algorithm consists of the time $T_{\mathrm{gen}}$ needed to generate $N$ random samples and the time $T_{\mathrm{arr}}$ needed to find the transformation $t^*$ whose $\delta$-neighborhood contains the most registered transformation vectors.

Assume that shapes are sets of polygons, without loss of generality, sets of triangles. 
A random point in a triangle can be generated in constant time using barycentric coordinates. For generating a random point from a set of $n$ triangles we  select a triangle randomly with probability proportional to the relative area of the triangle and then take a random point from the selected triangle. 
We first compute the areas of the triangles and partition the unit interval $[0,1]$ by subintervals whose lengths are proportional to these areas. Then the selection of a random $a\in [0,1]$ and a binary search on this partition gives us a random triangle.
Thus, we get preprocessing time linear in $n$ and $O(\log n)$ generation time for a single point.
Therefore, $T_{\mathrm{gen}} (n, N ) = O(n + N \log n)$.

Determining a translation whose $\d$-neighborhood obtained the most ``votes'' can be done by traversing the arrangement $\mathcal{A}$ given by the boundaries of the $\delta$-neighborhoods of the $N$  votes  from the random experiments.
The depth of a cell is defined as the number of neighborhoods it is contained in. The candidates for the output of the algorithm are the transformations contained in the deepest cells in this arrangement because a transformation $t$ lies in the intersection of $k$ of the neighborhoods if and only if its neighborhood contains $k$ votes. The size of the arrangement is $O(N^2)$  for translations and $O(N^3)$ for rigid motions.
The deepest cells can be determined by constructing and traversing the complete arrangement, which can be accomplished in time $O(N^2)$ for translations and $O(N^3)$ for rigid motions.

The runtime can be improved if, instead of the deepest cell in the arrangement,
an approximately deepest cell is computed.
If the depth of the arrangement is $d$, a witness point of depth $k$
such that $(1-\e) d \leq k \leq d$ can be computed in 
time $T_{\mathrm{arr}}(N)=O(N \e^{-2} \log N)$ \cite{approximateDepth}.
The total runtime of the algorithm is then $O(N\e^{-2}\log N + N \log n + n)$. We will show later  that the quality of the output can still be guaranteed if we approximate the depth.

In the following theorem we refer to the probabilistic algorithms as described in Section~\ref{sec:alg} except that in step 2 of each algorithm a transformation with an \emph{approximately} largest number of ``votes'' in its $\delta$-neighborhood is returned.
\begin{theorem}
   \label{thm:time}
   Let $A$ and $B$ be two shapes represented by sets of $n$ triangles in total. Let $\topt$ denote the transformation maximizing the area of overlap of $A$ and $B$ For a given error tolerance $\e$ and maximal allowed failure probability $\tau$ with $0< \e,\tau < 1$, the three algorithms described in Section~\ref{sec:alg} in combination with the  depth approximation algorithm of \cite{approximateDepth} compute a transformation $\tdpt$, such that 
\[
\left| \ \mt[\tdpt] - \mt[\topt] \ \right| < \e \mA
\]
with  probability at least $1-\tau$ in time 
$O(c/\e^8 \log (\max\left\{ 1/\tau,\ c/\e \right\}) \log (cn/\e) +n)$  for translations (algorithm \textsc{ProbMatchT})    and in time
$O(C/\e^{10} \log (\max\left\{ 1/\tau,\ C/\e \right\}) \log (Cn/\e) +n)$  for rigid motions with random rotation angle (algorithm \textsc{ProbMatchRMRA}).
Therein  $c=\mB^2 \Delta^4/\mA^4$, $C=\mB^2 \Delta^6 D^6/\mA^6$, 
$\Delta$ is  the length of the boundary of $A$, and $D$ is the  diameter of $A$.

For algorithm \textsc{ProbMatchRM3+1} the shapes $A$ and $B$ are additionally required to be $\kappa$-fat for some $0<\kappa\le 1$. The running time of the algorithm is then \\ $O\left( \max\left\{ \frac{C'}{\e^{10} \kappa^5}  \log \left( \max\left\{ \frac{1}{\tau},\ \frac{C'}{\e \kappa} \right\} \right), \ \frac{1}{\kappa^6} \log \frac{1}{\tau} \right\} \right)$, where  $C'=\mB^2 \Delta^6 D^6/\mA^6$.
\end{theorem}


\subsection{Overview of the Analysis}

In Section~\ref{sec:densityFunctions} we analyze the probability distribution implicitly given in the transformation space by the random experiment. It turns out that in the case of translations and in the case of rigid motions where the rotation angle is chosen randomly the density function  is proportional to the function mapping a transformation vector to the area of overlap of the transformed shape $A$ and $B$. For the rigid motions and algorithm \textsc{ProbMatchRM3+1} the density function is proportional to the squared value of the area of overlap.
Further, we prove that the density functions are Lipschitz continuous. Therefore, the probability of a $\delta$-neighborhood of a transformation $t$ converges uniformly to the value of the density function at $t$  times the size of the $\delta$-neighborhood as $\delta$ approaches zero.

Then in Section \ref{sec:absoluteError} we show that the relative number of transformations generated by random experiments that are contained in the $\delta$-neighborhood of a transformation $t$ is a good approximation of the probability of that $\delta$-neighborhood, in the sense that the probability of a large error decreases exponentially in the number of experiments.

Finally, we combine the uniform continuity of the density functions and the probability approximation results to derive rigorous bounds on the number of experiments required to find a transformation that with high probability approximates the maximum area of overlap within the given error bound.

\section{Density Functions}\label{sec:densityFunctions}

\subsection{Determining the Density Functions}

In this section we analyze the density functions of the probability distribution induced by the random experiments of the algorithm in the transformation space. We show that for translations and for rigid motions with random rotation angle the value of the density function for a transformation $t$  is proportional to the area of overlap $\mt$. For the algorithm \textsc{ProbMatchRM3+1} the induced density function is proportional to the squared area of overlap. Additionally, we show that in all three cases the density functions are Lipschitz continuous.

For deriving the density functions underlying the random experiments we will use the following probability theoretical transformation formula for density functions of random variables, see for example  \cite{krengel}.

\begin{theorem}
\label{thm:transformation_formula}
Let $X:\R^n \to \R^n$ be a random variable with density function $f_X$, open set $G\subset\R^n$ be the support of $f_X$
and $\varphi:G \to \R^n$ a continuously differentiable injective map, i.e., $\varphi:G\to G'$, where $G'=\varphi(G)$, is a bijection.
Let $\D (x) = \det (\frac{\partial \varphi_{i}}{\partial x_{j}}(x))_{i,j=1,\dots,n}$.
Then $\varphi \circ X$ has the density function
\[
f_{\varphi \circ X} (y) = 
\begin{cases}
   f_X(\varphi^{-1}(y))\ \vert \D(\varphi^{-1}(y)) \vert^{-1} & \text{for } y\in G' \\
   0 & \text{for }y\notin G'
\end{cases}
\enspace .
\]
\end{theorem}	

For translations and rigid motions with random rotation angle we can apply the following special case:

\begin{cor}
\label{thm:transformation_formula_affine}
Let $X:\R^n \to \R^n$ be a random variable with density function $f_X$
and $h:\R^n \to \R^n, h: x \mapsto Mx$ a linear map with $\det(M) \neq 0$.
Then $h \circ X$ has the density function
\[
f_{h \circ X} (y) = f_X(M^{-1}y)\ \vert \det(M^{-1}) \vert.
\]
\end{cor}

For a subset $B$ of a set $R$ let $\chi_B:R \to \{ 0,1 \}$ be the
characteristic function of $B$ that is 1 if a point from $R$ is in $B$
and 0 otherwise.

\paragraph{The Density Function for Translations. }

\begin{lemma}[Translations]\label{la:density}
 The density function of the probability distribution
	on the translation space that results from the experiment in algorithm  \textsc{ProbMatchTrans} is given by
	\[
	g_T(t) = \mt / (\mA \mB) \enspace .
	\]
\end{lemma}

\begin{proof}
We model the experiment by regarding $X=id_{A \times B}$ on $\R^4$
as uniformly distributed random variable. $X$ corresponds to the sample pairs selected by the random experiment. The density function of $X$ is 
\[
f_X (a,b) = \frac{\chi_A(a) \ \chi_B(b)}{\mA \mB} \enspace .
\]
Consider the bijective function $\varphi:\R^4 \to \R^4, \varphi:(a,b) \mapsto (a,b-a)$. $\varphi$ maps a pair of points $(a, b)$ to a point-translation pair $(a,t)$ where $t$ is the translation that maps $a$ to $b$. 
By Corollary \ref{thm:transformation_formula_affine} the density function of $\varphi \circ X$ is
\[
f_{\varphi \circ X} (a,t) = \frac{\chi_A(a) \ \chi_B(a+t)}{\mA \mB} 
= \frac{\chi_{A \cap (B-t)}(a)}{\mA \mB} \enspace .
\]
The density function on the translation space $\R^2$ is the density function of the projection of $\varphi \circ X$ to the last two coordinates: 
\[
g_T(t) = \int_{a\in \R^2} f_{\varphi \circ X} (a,t) da 
= \frac{\vert A \cap (B-t) \vert}{\mA \mB}
= \frac{\mt}{\mA \mB} \enspace . \qedhere
\]
\end{proof}

\paragraph{The Density Function for Rigid Motions with Random Angle. }

\begin{lemma}[Rigid Motions with Random Angle]
   \label{la:rmra}
The density function on the space of rigid motions~$R$ induced by algorithm  \textsc{ProbMatchRMRA} is given by
	\[
	g_{RA}(r) = \mt[r] /(\mA \mB).
	\]
\end{lemma}

\begin{proof}
Our random experiment consists in selecting uniformly distributed
points from $\Omega=I\times A\times B$ where $I=[-1/2,1/2)$.
We are interested in the density function $f_Y$ of the random variable
\[
Y: \Omega \to R, \quad Y: (\a,a,b) \mapsto (\a, b-M_\a a)\enspace .
\]
We will express the density function of $Y$ in terms of the conditional probability densities of the following two random variables $Y_I$ and $Y_T$ defined as
\begin{align*}
   Y_I: \Omega &\to I, &  Y_I: (\a,a,b) &\mapsto \a \enspace, \\
   Y_T: \Omega &\to \R^2, &  Y_T: (\a,a,b) &\mapsto b-M_\a a \enspace.
\end{align*}
The density function of $Y$ is the joint density of the random variables $Y_I$ and $Y_T$. Recall that the  counterclockwise rotation angle is selected uniformly distributed in $I$ independently from the points $a$ and $b$. So the marginal probability density of $Y_I$, i.e., probability density of $Y_I=\a$ allowing all possible values of $Y_T$, is \[f_I(\a)=\genfrac{}{}{}{1}{1}{\vert I \vert}=1\enspace .\]
The value of $Y_T$ depends on the selected points $a$ and $b$ and on the value of $Y_I$. The conditional probability density of $Y_T=t$ given $Y_I=\a$ is exactly the probability density in the space of translations for shapes $M_\a A$ and $B$:
\[ f_T(t \ | \ Y_I=\a ) = \frac{| (M_\a A + t)\cap B |}{|A| |B|} \enspace. \]
The conditional probability density can also be expressed in terms of the joint probability density
$f_T(t \ | \ Y_I=\a ) = f_Y(\a,t) / f_I(\a)$.
Thus we get for any rigid motion $r=(\a,t)$ that
	\[
	g_{RA}(r) = f_Y(r) = \mt[r] /(\mA \mB) \enspace . \qedhere
	\]
\end{proof}

\paragraph{The Density Function for Rigid Motions with 3+1 Points. }

\begin{lemma}[Rigid Motions with 3+1 Points]
   \label{la:rm31}
The density function on the space of rigid motions~$R$ induced by the algorithm  \textsc{ProbMatchRM3+1} is given~by
	\[
	g_{3+1}(r) = \vert r(A) \cap B \vert^2/c \enspace ,
	\]
where $c$ is a positive real depending on $A$ and $B$ which is at most $\vert A \vert^2 \vert B \vert$.
\end{lemma}

\begin{proof}
In one random experiment we select uniformly distributed random elements from the set
\[
\mathcal{S}=\{ (a_{1},a_{2},b_{1},\beta) \in A^2 \times B \times I : a_{1} \neq a_{2},
b_{2}=b_{1}+ \Vert a_{2}-a_{1}\Vert M_{\beta} \genfrac{(}{)}{0pt}{1}{1}{0} \in B \}\enspace .
\] 
The density function of the random variable $X=id_{\mathcal{S}}$ is then
\begin{align*}
   f_X(a_1, a_2, b_1, \beta) &=\frac{\chi_{\mathcal{S}}(a_1,a_2,b_1,\beta)}{|\mathcal{S}|}\\
   &= \frac{\chi_{A}(a_{1}) \chi_{A}(a_{2}) \chi_{B}(b_{1}) \chi_{B} (b_{2})}{|\mathcal{S}|} \enspace ,
\end{align*}
where $b_2$ is as in the definition of $\mathcal{S}$.

Let $Y$ denote the random variable corresponding to the rigid motion resulting from one random experiment:
\[ Y:\mathcal{S}\to R\enspace , \quad Y: (a_{1},a_{2},b_{1},\beta) \mapsto (\a,b_{1}-M_{\a}a_{1}) \enspace , \]
where $\alpha= \angle (a_{2}-a_{1},b_{2}-b_{1})/2\pi$.
We represent $Y$ as a composite function of random variable $X$. Define functions $\varphi_1,\varphi_2$ as follows:
\begin{align*}
   \varphi_1: &\ \mathcal{S} \to  \R^6\times I &
\varphi_1: &\ (a_{1},a_{2},b_{1},\beta) \mapsto  (a_{1},a_{2},b_{1}-M_{\a}a_{1},\a) \\
\varphi_2: &\ \R^6\times I \to R &
\varphi_2: &\ (a_1,a_2,t,\a) \mapsto (\a,t) \enspace .
\end{align*}
Then $Y=\varphi_2\circ \varphi_1\circ X$. 

Observe that the set $\mathcal{S}$ is open, since the sets $A$ and $B$ are open, the excluded set of tuples where $a_1=a_2$ is a closed set, and the interval $I$ is equivalent to a unit circle and is therefore open. 
Further, the function $\varphi_1$ and its inverse are bijective and differentiable, so we can apply Theorem \ref{thm:transformation_formula} to $\varphi_1\circ X$.
We first compute the determinant $\Delta(x)$ of the Jacobian matrix of $\varphi_1$. 
It is easy to see that $\det (\frac{\partial \varphi_{1,i}}{\partial x_{j}})_{i,j=1,\dots,7}=
\det (\frac{\partial \varphi_{1,i}}{\partial x_{j}})_{i,j=5,\dots,7}$ since $\varphi_{1,i}(x_1,\dots,x_7)=x_i$ for $i=1,\dots,4$.
Note that the angle $\a$ does not depend on $b_1$ and it depends linearly on $\beta$: $\a=\beta-\gamma$, where  $\gamma= \angle(a_{2}-a_{1},\left( \substack{1\\0} \right)) / 2\pi$. Therefore, 
$\frac{\partial (b_{1} - M_{\a} a_{1})_{i} }{\partial b_{1j}} = \d_{ij}$,
$\frac{\partial \a }{\partial b_{1j}} = 0$ for $j=1,2$, 
and $\frac{\partial \a}{\partial \beta} =1$.
Now we have that
\[\left(\frac{\partial \varphi_{i}}{\partial x_{j}}\right)_{i,j=5,\dots,7}=\begin{pmatrix}1 & 0 & \lambda_1 \\ 0 & 1 & \lambda_2 \\ 0 & 0 & 1 \end{pmatrix} \enspace ,\]
for some $\lambda_1,\lambda_2\in \R$. Thus, $\D(x)=1$.
The inverse function of $\varphi_1$ maps a tuple $(a_{1},a_{2},t,\a)$ to $(a_{1},a_{2},t+M_{\a}a_{1}, \a + \gamma)$.
By Theorem \ref{thm:transformation_formula} we get that the density function of $\varphi_1\circ X$ is
\begin{align*}
f_{\varphi \circ id_{\mathcal{S}}}(a_{1},a_{2},t,\a)& = 
\chi_{A}(a_{1}) \chi_{A}(a_{2}) \chi_{B}(t+M_{\a}a_{1}) \chi_{B}(t+M_{\a}a_{2})
/\vert \mathcal{S} \vert\\ 
&= \chi_{A}(a_{1}) \chi_{A}(a_{2}) \chi_{M_{-\a}(B-t)}(a_{1}) \chi_{M_{-\a}(B-t)}(a_{2})
/\vert \mathcal{S} \vert \enspace .
\end{align*}

The density function of random variable $Y$ on $R$ is then
\begin{align*}
   g_{3+1}(\a,t) &=  1/\vert \mathcal{S} \vert \int_{A^2} \chi_{(A \cap r^{-1}(B))^2}(a_{1},a_{2}) d(a_{1},a_{2})\\ 
   & =  \vert A \cap r^{-1}(B)\vert^2 / \vert \mathcal{S} \vert \\
&= \vert r(A) \cap B \vert^2/\vert \mathcal{S} \vert \enspace ,
\end{align*}
where $r=(\a,t)$.
\end{proof}

\subsection{Lipschitz Continuity of the Density Functions.}
\label{sec:lipschitz}

In this section we show that the density functions of the probability distribution induced in the space of transformations by the algorithm are Lipschitz continuous.

A function $h$ from a metric space $M$ to $\R$ is called Lipschitz continuous if there is a constant $L$ such that for all $x,y \in M$ holds
\[ \Vert x-y \Vert < \d \Longrightarrow \vert h(x)-h(y) \vert < L \d \enspace .\]

Let $\dBall$ denote the $\delta$-neighborhood of a transformation $t$ with respect to the maximum norm and let $\mu_{\delta}$ be the Lebesgue measure of $\dBall$. 
We are interested in the density functions to be Lipschitz continuous because then $\prob/\md$ converges to $h(t)$ for $\d \to 0$ uniformly on the transformation space, where $P$ is the probability distribution with density function $h$.

\begin{lemma}
   For fixed shapes $A$ and $B$
   let $h$ be the density function of the probability distribution $P$ in the transformation space induced by the probabilistic algorithm. There exists a constant $L$ such that for every transformation~$t$
\[ \left| \prob - \md  h(t) \right| \leq L \md \d \enspace , \]
where $\dBall$ denotes the $\delta$-neighborhood of a transformation $t$ with respect to the maximum norm and  $\mu_{\delta}$ is the Lebesgue measure of $\dBall$.

For translations $L=L_T=\sqrt{2} \D/(\mA\mB)$, for rigid motions and algorithm \textsc{ProbMatchRMRA} $L=L_{RA}=(\sqrt{2}+2\pi D)\D/(\mA \mB)$, and for rigid motions and algorithm \textsc{ProbMatchRM3+1} $L=L_{3+1}= 2(\sqrt{2}+2\pi D)\D \min(\mA, \mB) /c $, where $D$ is the diameter and $\Delta$ the boundary length of $A$, and $c$ is the constant from Lemma~\ref{la:rm31}.
   \label{la:Lipschitz}
\end{lemma}
\begin{proof}
   Assume that $h$ is Lipschitz continuous with constant $L$ then
   \begin{align*}
      \prob=\int_{\dBall} h(x) d x & \le \md \sup_{x\in \dBall} h(x) \le  \md (h(t)+L\delta) \quad \text{ and }\\
      \int_{\dBall} h(x) d x & \ge \md \inf_{x\in \dBall} h(x) \ge \md (h(t)-L\delta) \enspace .
   \end{align*}

   It remains to show that for every variant of the probabilistic algorithm the induced probability density function is Lipschitz continuous.
Let $f(r)$ denote the area of overlap $r(A)\cap B$ for a rigid motion $r$. We first show that the function $f$ is Lipschitz continuous.
We assume without loss of generality that the input shapes $A$ and $B$ contain the origin.

Let $r=(\a,p),s=(\beta,q)$ be rigid motions whose distance is less than $\d$
in the maximum norm. Then the distance between translation vectors is $||p-q||\le \sqrt{2}\delta$ and $|\alpha -\beta|\le \delta$.

The difference in the area of overlap for $r$ and $s$ can be bounded by the area of the symmetric difference between $r(A)$ and $s(A)$:
\begin{align*}
   f(r)=|r(A) \cap B| & = |(r(A)\setminus s(A))\cap B| + |(r(A)\cap s(A))\cap B| \\
   & \le |r(A)\setminus s(A)| + |s(A)\cap B| \enspace .
\end{align*}
Thus, $f(r)-f(s) \le |r(A)\setminus s(A)|$. Similarly, $f(s)-f(r) \le |s(A)\setminus r(A)|$. Combining these two estimates we get 
\begin{align*}
   |\ f(r)-f(s) \ | & \le  \max\left( |r(A)\setminus s(A)|,\ |s(A)\setminus r(A)| \right) \enspace . 
\end{align*}
Let $w(\d)$ be the maximal length of the line segment $\overrightarrow{r(x)s(x)}$ for $x \in A$.
The difference in the area of overlap  is at most $ \D w(\d)$ since $r(A)$ and $s(A)$  differ by at most a $w(\d)$-wide strip along the boundary of $r(A)$. Next we find an upper  bound on the length $w(\delta)$.

By an easy geometric argument the distance between a point  $x$ and its rotated image can be expressed as $|| x-M_{\alpha} x || = 2||x|| \cdot |\sin (2\pi \alpha / 2)|$ for $0\le |2\pi \alpha| \le \pi /2$. Since $|\sin \gamma |\le |\gamma|$ for all $\gamma$ this distance can be bounded by $|| x-M_{\alpha} x || \le ||x|| \cdot |2\pi \alpha|$.

Let $x\in A$ then $|| x || \le D$. We use the above argument to bound the distance between the image of $x$ under rigid motions $r$ and $s$. Observe that since $\delta$ is small we can assume that $0\le |2\pi \delta| \le \pi /2$.
\begin{align*}
   || r(x) - s(x) || & = || M_{\alpha} x+ p -M_{\beta}x - q || \\
   & \le || M_{\alpha} x -M_{\beta}x || + || p - q || \\
   & \le || M_{\alpha - \beta} x || + \sqrt{2}\delta \\
   & \le 2\pi | \alpha -\beta |\cdot  ||x|| +\sqrt{2}\delta \\
   & \le 2\pi\delta D + \sqrt{2}\delta = (\sqrt{2}+2\pi D)\delta 
\end{align*}
for all $x$ in $A$.

Then  for arbitrary rigid motions $r$ and $s$ such that $\Vert r-s \Vert < \d$ the difference in the area of overlap can be bounded by $\vert f(r)-f(s) \vert < (\sqrt{2}+2\pi D)\d \Delta $. The Lipschitz constant for the area of overlap is then $L_f= (\sqrt{2}+2\pi D) \Delta $.

The density function of the probability distribution in the space of rigid motions induced by the algorithm \textsc{ProbMatchRMRA} is $g_{RA}(r)=f(r)/(\mA \mB)$ by Lemma~\ref{la:rmra}. Then for rigid motions $r$ and $s$ such that $\Vert r-s \Vert < \d$ we get
\[
| g_{RA}(r)-g_{RA}(s) | = \vert f(r)-f(s) \vert / (\mA \mB) < (\sqrt{2}+2\pi D) \Delta \delta / (\mA \mB) \enspace .
\]
Thus, the Lipschitz constant of the function $g_{RA}$ is $L_{RA}=(\sqrt{2}+2\pi D)\D/(\mA \mB)$.

The density function  induced by the algorithm \textsc{ProbMatchRM3+1} is $g_{3+1}(r)=f^2(r)/c$, where $c$ is the constant from Lemma~\ref{la:rm31}. 

Observe that if $f$ is bounded and Lipschitz continuous with Lipschitz constant $L_f$, then $f^2$ is also Lipschitz continuous with  constant $L_{f^2}=2L_f \sup_x f(x)$ due to the following consideration:
\begin{align*}
   | f^2(x) -f^2(y) | &= | f(x) -f(y) |\cdot | f(x)+f(y) | < L_f \cdot | x-y| \cdot | f(x) + f(y) | \\
   &\le 2 L_f | x-y| \sup_z f(z) \\
   & =2(\sqrt{2}+2\pi D)\D  | x-y| \sup_z f(z)  /c \enspace .
\end{align*}
Thus, the Lipschitz constant of the function $g_{3+1}$ is 
\begin{align}
   L_{3+1}=2(\sqrt{2}+2\pi D)\D  \sup_z f(z)  /c \enspace .
   \label{eq:l31}
\end{align}
The maximal possible area of overlap of two shapes under rigid motions is clearly bounded by the  area of the smaller shape. Therefore, the function $g_{3+1}$ is Lipschitz continuous with constant $L_{3+1}\le 2(\sqrt{2}+2\pi D)\D \min(\mA, \mB) /c $.

In the case of translations we can disregard rotation, so the Lipschitz constant $L_{RA}=(\sqrt{2}+2\pi D)\D/(\mA \mB)$ for the density function in the case of rigid motions reduces to
$L_T=\sqrt{2} \D/(\mA\mB)$ for translations.

Note that the constants depend heavily on the shapes.
\end{proof}

\section{Absolute Error Approximation }\label{sec:absoluteError}
In the previous section (Lemma~\ref{la:Lipschitz}) we showed that for the probability distributions in the space of transformations induced by the algorithms for translations and for rigid motions the value of the probability function $P$ for a $\delta$-neighborhood $\dBall$ (divided by the measure $\md$ of the $\delta$-neighborhood) of a transformation $t$ converges to the value of the density function $h(t)$ for that transformation as $\delta$ approaches zero.   

In this section we prove that the relative number  of ``votes'' in the $\delta$-neigh\-bor\-hood of a transformation $t$ is a good approximation of the probability $P(\dBall)$ and complete the proofs of Theorem~\ref{thm:main_absolute} and Theorem~\ref{thm:time}. 

\comment{
\paragraph{Chernoff bound thm}
}
Let the random variable $X_N^{\delta}(t)$ denote the number of registered transformations  in the $\delta$-neighborhood of transformation $t$.
We use the Chernoff bound formulated as in \cite{finding_a_guard}
for proving that for fixed $t$ with high probability 
$X_N^\d(t)/N$ and $\prob$ do not differ much for large $N$.
\begin{theorem}[\cite{finding_a_guard} Chernoff bound]
Let $X_1,\ldots,X_N$ be independent binary random variables,
let $X=\sum_{i=1}^NX_i$, and let $0<\epsilon<1$. Then
\[
P(\vert X - E(X) \vert > \e N) < 2e^{-\e^2N/2}.
\]
\label{thm:chernoff}
\end{theorem}
Applying this bound to our setting yields
\begin{cor}\label{cor:chernoff}
For each transformation $t$ and for all $0 < \e < 1$
\[
P(\vert X_N^\d(t)/N - \prob \vert > \e) < 2e^{-\e^2N/2}.
\]
\end{cor}
\begin{proof}
Define $X_i=\chi_{\dBall}$ for $i=1,\ldots,N$.
The $X_i$ are identically distributed, independent, binary random variables
with $X_N^\d(t)=\sum_{i=1}^NX_i$ and $E(X_N^\d(t))=N\prob$.
The inequality of Corollary~\ref{cor:chernoff} results from applying the Chernoff bound to $X_1,\ldots,X_N$.
\end{proof}
This shows that $X_N^\d(t)/N$ converges to $\prob$ in probability
as the number of random experiments goes to infinity for each $t$
in the transformation space.

We have already seen that for fixed $t$
\[
X_N^\d(t)/N \quad 
\substack{N \to \infty \\ \longrightarrow \\ \text{in prob.}}
\quad \prob \quad 
\substack{\d \to 0 \\ \longrightarrow \\ \text{uniformly}}
 \quad \md \ h(t) \enspace .
\]
Now we need to analyze what happens if the transformation vector
is determined by the sequence of random experiments,
namely a vector whose $\d$-neighborhood obtains the most ``votes'',
and thus is a random vector itself.

\comment{
\paragraph{Key lemma}
}
The output of the algorithm can be modeled as a random variable
\[
Z_N^\d=\max_{t} X_N^\d(t).
\]
Let $S=(s_1,\ldots,s_N)$ be a sequence of transformations from the random experiments.
Consider the arrangement $\mathcal{A}$ induced by the boundaries of 
$\dBall[s_1],\ldots,\dBall[s_N]$, which are the $\d$-spheres with respect to the maximum norm of the points in $S$.
The depth of a cell is defined as the number of $\dBall[s_i]$
it is contained in. The candidates for the output of the algorithm
are the transformations contained in the deepest cells in this arrangement.
A transformation $t$ lies in the intersection of $k$ of the neighborhoods
if and only if its neighborhood contains $k$ ``votes''.

The next lemma can be proven using an idea of \cite{finding_a_guard}.

\begin{lemma}[Key Lemma]
   \label{la:key}
Let $V$ be the set of all vertices of the arrangement~$\mathcal{A}$ and $\topt$ the transformation maximizing the area of overlap $|t(A)\cap B|$.
Then for each $\e>0$ and $N > 6/\e +2$ holds in the case of translations
\[
P(\exists t \in V \cup \{ \topt \} : \vert X_N^\d(t)/N - \prob \vert > \e)
< 2N^2e^{-\e^2(N-2)/8} \enspace ;
\]
and in the case of rigid motions,
\[
P(\exists t \in V \cup \{ \topt \} : \vert X_N^\d(t)/N - \prob \vert > \e)
< \frac{2}{3}N^3 e^{-\e^2(N-3)/8} \enspace .
\]
\end{lemma}
\begin{proof}
In the case of translations, a vertex of $\mathcal{A}$ is defined by 2 $\d$-spheres, and 2 $\d$-spheres define at most 2 vertices, so $\vert V \vert \leq N(N-1)$ and $\vert V \cup \{\topt\} \vert \leq N^2$.
In the case of rigid motions, a vertex of $\mathcal{A}$ is defined by 3 $\d$-spheres, and 3 $\d$-spheres define at most 2 vertices, so $\vert V \vert \leq 2 {N \choose 3} = N^3/3 - N^2 + 2N/3$ and $\vert V \cup \{\topt\} \vert \leq N^3/3$.

Let us consider translations first. Let $t$ be a vertex of $\mathcal{A}$, then $t$ is an intersection point of two $\delta$-spheres $\dBall[v], \dBall[w]$ for some votes $v,w$. Consider same random sequence  without two experiments yielding $v$ and $w$.
Since the random points are independent and the choice of $t$ does not depend on the shorter random sequence, we can apply Corollary \ref{cor:chernoff} to $t$ with $N-2$ experiments.
 \begin{align*}
P ( \vert X_{N-2}^\d(t)/(N-2) - \prob \vert > \e/2 )
& <  2e^{-\e^2(N-2)/8} \enspace .
\end{align*}

For each $\d$-neighborhood, the fractions of ``votes'' that lie in the neighborhood differ by at most $2/(N-2)$ if two arbitrary ``votes'' are deleted.
Therefore, 
\begin{align*}
   |X_N^{\delta}(t)/N -\prob|& \le |X_{N-2}^{\delta}(t)/(N-2) - \prob|+2/(N-2) \enspace.
\end{align*}
Choosing  $N \geq 4/\e +2$ we get that $2/(N-2) \leq \e/2$ and
\begin{align*}
   P ( \vert X_{N}^\d(t)/(N) - \prob \vert > \e ) & <  2e^{-\e^2(N-2)/8} \enspace .
\end{align*}
As argued above there are at most $N^2$ such vertex points together with the point $\topt$. Applying the triangle inequality we get the claim of the Lemma for the case of translation.

The argumentation for rigid motions is analogous.
\end{proof}

\comment{
\paragraph{cor:output}
}
Next we show that if the probability density function $h$ is Lipschitz continuous  then the value of $\prob$ for two transformations that lie in one $\delta$-neighborhood does not differ  much.
\begin{prop}
   \label{pro:cell}
Let $r,s,t$ be transformations such that $r,s \in \dBall$.
Then \linebreak $\vert \prob[r] - \prob[s] \vert < 4L\md \d$ where $L$
is the Lipschitz constant of the density function $h$ of $P$ and $\mu_{\delta}$ is the Lebesgue measure of a $\dBall$.
\end{prop}
\begin{proof}
   \begin{align*}
      \vert \prob[r] - \prob[s] \vert &=
\left| \int_{\dBall[r]} h(v)dv - \int_{\dBall[s]} h(v)dv \right|
\leq \md (\sup_{\dBall[r]} h(v) - \inf_{\dBall[s]} h(v)) \\
&< 4L\md \d \enspace . \qedhere
   \end{align*}
\end{proof}

It follows from Lemma~\ref{la:key} and  Proposition~\ref{pro:cell} that  with high probability  for every transformation $t$ the relative number of votes in its $\delta$-neighborhood is a good approximation of the probability~$\prob$.
\begin{cor}\label{cor:output}
   Let $0<\e<1$ and  $N \geq 6/\e +2$. For all transformations $t$  with probability at least $1-2N^2e^{-\e^2(N-2)/8}$ in the case of translations and with probability
at least $1-\frac{2}{3}N^3 e^{-\e^2(N-3)/8}$ in the case of rigid motions it holds
\[
\vert X_N^\d(t)/N - \prob \vert < \e + 4L\md \d \enspace ,
\]
where $L$ is the Lipschitz constant of the probability density function.
\end{cor}
\begin{proof}
   Let $t$ be an arbitrary transformation and let $t'$ be a vertex of the cell in $\mathcal{A}$ which contains $t$, such that $X_N^{\delta}(t')=X_N^{\delta}(t)$. Then 
   \begin{align*}
      |X_N^{\delta}(t)-\prob| & \le |X_N^{\delta}(t')-\prob[t']| + |\prob[t']-\prob| \\
      & \le \e + 4 L\delta\md \qquad \text{by Lemma~\ref{la:key} and Proposition~\ref{pro:cell}}\enspace . \qedhere
   \end{align*}
\end{proof}
That means that the output of the algorithm does not need to be chosen among the vertices of the arrangement $\mathcal{A}$.

\comment{
\paragraph{la:hApprox}
}
Next we show that if the probability density function is Lipschitz continuous then the transformation with the maximum number of ``votes'' in its $\delta$-neighborhood results in a good approximation of the maximum of the density function with high probability.
\begin{lemma}
   Let $h(t)$ be the density of the probability distribution $P$ in the transformation space induced by the algorithm, and $L$ the Lipschitz constant of $h$. Let $\topt$ be the transformation maximizing $h(t)$ and $\tapp$ the transformation maximizing $X_N^{\delta}(t)$, where $X_N^{\delta}(t)$ denotes the number of transformations generated by the random experiments that are contained in the $\delta$-neighborhood of~$t$. The Lebesgue measure of a $\delta$-neighborhood is denoted by $\md$. Then for all $0<\e\le 1$
   \[ h(\tapp) \ge h(\topt) - 2\e/\md -6 L\delta \enspace \]
   with probability at least $1-q$, where $q=2N^2e^{-\e^2(N-2)/8}$ for translations and  $q=\frac{2}{3}N^3 e^{-\e^2(N-3)/8}$ in the case of rigid motions.
   \label{la:hApprox}
\end{lemma}
\begin{proof}
It follows from Lemmas \ref{la:Lipschitz} and \ref{la:key} and from Corollary~\ref{cor:output} that with probability at least $1-q$ the following two statements are true:
\begin{align*}
   |\md h(\tapp)-X_N^{\delta}(\tapp)/N| & \le |\md h(\tapp) -P(\dBall[\tapp])| + |P(\dBall[\tapp])-X_N^{\delta}(\tapp)/N| \\
   &\le L\delta\md + \e + 4L\md \delta \qquad \text{by Lemma~\ref{la:Lipschitz} and Corollary \ref{cor:output}} \\
   &\le \e+5L\md\delta 
   \intertext{and}
   \vert \md h(\topt) - X_N^\d(\topt)/N \vert & \le |\md h(\topt) -P(\dBall[\topt])| + |P(\dBall[\topt])-X_N^{\delta}(\topt)/N| \\
   &\leq  L \md \d +\e  \qquad \text{by Lemmas~\ref{la:Lipschitz} and \ref{la:key}}\enspace . 
\end{align*}
By the definition of $\tapp$,  $X_N^{\delta}(\tapp)\ge X_N^{\delta}(\topt)$. Then 
\begin{align}
   h(\tapp) & \ge (X_N^{\delta}(\tapp)-\e -5L\md \delta) / \md \notag \\
   & \ge (X_N^{\delta}(\topt)-\e -5L\md \delta) / \md \notag \\
   & \ge (\md h(\topt)-\e - L\md \delta -\e -5L\md \delta) / \md \notag \\
   & = h(\topt) - 2\e / \md -6L \delta \enspace , \notag 
\end{align}
which concludes the proof.
\end{proof}

\comment{
\paragraph{proof of thm 1}
}
Now we can prove Theorem \ref{thm:main_absolute}.
\begin{proof}[Proof of Theorem \ref{thm:main_absolute}]
   Let $f(t)$ denote the area of overlap $t(A)\cap B$, $h(t)$ the density of the probability distribution $P$ in the transformation space induced by the algorithm, and $L$ the Lipschitz constant of $h$. We will use two parameters $0< \delta, \eta \le 1$, which will be determined later. The parameter $\delta$ specifies the size of the neighborhoods used by the matching algorithms, and $\eta$ is the error tolerance value used when applying lemmas or corollaries. 

Recall that $\tapp$ is a transformation with the maximum number of transformations generated by the random experiments in its $\delta$-neighborhood and $\topt$ the transformation maximizing $f(t)$. Let $q$ denote the failure probability from Lemma~\ref{la:key} in terms of the number of experiments $N$ and the error tolerance $\eta$, for translations $q=2N^2e^{-\eta^2(N-2)/8}$ and for rigid motions $q=\frac{2}{3}N^3 e^{-\eta^2(N-3)/8}$.

Since $\mt[\tapp] \leq \mt[\topt]$ holds by definiton,
we only have to show that $\mt[\tapp] \geq \mt[\topt] - \e \mA$. By Lemma~\ref{la:hApprox} for translations and rigid motions 
\begin{align}
   h(\tapp)\ge h(\topt) - 2\eta / \md -6L \delta
   \label{eq:la16}
\end{align}
with probability at least $1-q$.

\smallskip\noindent
\emph{Translations: } In the case of translations we have by Lemma~\ref{la:density} $h(t)=\mt/\mA\mB=f(t)/\mA\mB$, $L=\sqrt{2}\Delta/\mA\mB$ by Lemma~\ref{la:Lipschitz}, and $\md=4\delta^2$. Then by \eqref{eq:la16}
\begin{align*}
   f(\topt)-f(\tapp) & \le \frac{2\eta \mA\mB}{4\delta^2}+6\sqrt{2}\delta\Delta \enspace.
\end{align*}
Bounding this error to be at most $\e\mA$, we get 
\begin{align*}
   \eta & \le \frac{4\delta^2\e\mA}{2\mA\mB} - \frac{24\sqrt{2}\delta^3\Delta}{2\mA\mB} \enspace .
\end{align*}
In order to maximize this expression we differentiate it with respect to $\delta$ and determine the value of $\delta$ for which that derivative is zero.
The value of $\delta$ maximizing the above expression and the corresponding value of $\eta$ are then
\begin{align}
   \delta & = \frac{\e\mA}{9\sqrt{2}\Delta} \qquad \text{ and } \qquad \eta  = \frac{\e^3 \mA^2}{243\Delta^2\mB} \enspace . \label{eq:dht}
\end{align}
Finally, we want the probaility of failure $q=2N^2e^{-\eta^2(N-2)/8}$ to be at most $\tau$. A straightforward analysis shows that for $N \geq 80/\eta^2 \log (80/\eta^2)$ this value is at most $ e^{-\eta^2(N-2)/16}$.
Solving the inequality $ e^{-\eta^2(N-2)/16}\le \tau$ with respect to $N$ we get that for
$N \geq \max \{ - 16/\eta^2 \log(1/ \tau) + 2,\ 80/\eta^2 \log (80/\eta^2) \}$ the approximation error is at most $\e \mA$ with probability at least $1-\tau$.

\smallskip\noindent
\emph{Rigid motions with random rotation angle: } Here we have by Lemma~\ref{la:rmra} $h(t)=\mt/\mA\mB=f(t)/\mA\mB$, $L=(\sqrt{2}+2\pi D)\Delta/\mA\mB$ by Lemma~\ref{la:Lipschitz}, and $\md=8\delta^3$. Then by \eqref{eq:la16}
\begin{align*}
   f(\topt)-f(\tdpt) & \le \frac{2\eta \mA\mB}{8\delta^3}+6(\sqrt{2}+2\pi D)\Delta\delta \enspace.
\end{align*}
Again, bounding this error by $\e\mA$ and maximizing the expression for $\eta$ as in the case of translations we get that the optimal choice for $\delta$ and $\eta$ is
\begin{align}
   \delta & = \frac{\e\mA}{8(\sqrt{2}+2\pi D)\Delta}  \qquad \text{ and } \qquad \eta  = \frac{\e^4 \mA^3}{512(\sqrt{2}+2\pi D)^3\Delta^3\mB} \enspace . \label{eq:dhrm}
\end{align}
By a straightforward analysis the probability of failure $q=\frac{2}{3}N^3 e^{-\eta^2(N-3)/8}$ is at most $ e^{-\eta^2(N-2)/16}$ for all $N\ge 112/\eta^2 \log (112/\eta^2)$. Solving the inequality  $e^{-\eta^2(N-2)/16}\le \tau$ with respect to $N$ we get that for all $N \geq \max \{ - 16  /\eta^2 \log (1/\tau) + 3, 112/\eta^2 \log (112/\eta^2) \}$ the approximation error of the algorithm is at most $\e\mA$ with probability at least $1-\tau$.
\end{proof}

\comment{
\paragraph{proof thm 2}
}
Next we prove Theorem~\ref{thm:abs31} in a similar fashion.
\begin{proof}[Proof of Theorem \ref{thm:abs31}]
   Since in algorithm \textsc{ProbMatchRM3+1} some random samples are rejected and, therefore, do not induce a vote in the transformation space, we first determine the necessary number of not rejected experiments, in the following denoted by $M$, in order to guarantee the required error bound with high probability. Afterwards, we determine the total number $N$ of random samples that the algorithm needs to generate in order to record at least $M$ votes in the transformation space with high probability.

   Let $f(t)$ denote the area of overlap $t(A)\cap B$, $h(t)$ the density of the probability distribution $P$ in the  space of rigid motions induced by  algorithm \textsc{ProbMatchRM3+1}, and $L$ the Lipschitz constant of $h$. 

Let $\tapp$ be a transformation with the maximum number of transformations generated by the  random experiments in its $\delta$-neighborhood and $\topt$ the transformation maximizing $f(t)$. Let $q$ denote the failure probability from Lemma~\ref{la:key} in terms of the number of votes $M$ and the error tolerance~$\eta$,  $q=\frac{2}{3}M^3 e^{-\eta^2(M-3)/8}$.

Since $\mt[\tapp] \leq \mt[\topt]$ holds by definition,
we only have to show that $\mt[\tapp] \geq \mt[\topt] - \e \mA$. By Lemma~\ref{la:hApprox}  $h(\tapp)\ge h(\topt) - 2\eta / \md -6L \delta$ with probability at least $1-q$.

For algorithm \textsc{ProbMatchRM3+1} the density function $h(t)=f^2(t)/c$, where $c$ is the constant from Lemma~\ref{la:rm31}. The Lipschitz constant is $L=2(\sqrt{2}+2\pi D)\D \min(\mA, \mB) /c$ by Lemma~\ref{la:Lipschitz}, where $D$ is the diameter and $\Delta$ the length of the boundary of $A$,  and the size of the $\delta$-neighborhood is $\md=8\delta^3$. In the proof of Lemma~\ref{la:Lipschitz} we actually showed a stronger bound on the Lipschitz constant (Equation~\eqref{eq:l31}), namely $L_{3+1}=2(\sqrt{2}+2\pi D)\D  \sup_z f(z)  /c$, which is $2(\sqrt{2}+2\pi D)\D   f(\topt)  /c$. This stronger bound is used in the computations below. 

Additionally, since the shapes $A$ and $B$ are $\kappa$-fat, and the shape $A$ is the one with the smaller largest inscribed circle, the maximal area of overlap of $A$ and $B$ under rigid motions is at least as large as the area of the largest inscribed circle in $A$, that is, $f(\topt)\ge \kappa\mA$.

By Lemma~\ref{la:hApprox} we get that with probability at least $1-q$
\begin{align*}
   \frac{1}{c}(f(\topt)-f(\tapp))(f(\topt)+f(\tapp))& = \frac{f(\topt)^2}{c} -\frac{f(\tapp)^2}{c} \\
   & = h(\topt) - h(\tapp) \\& \le \frac{2\eta}{\md} - 6L\delta \\
   & = \frac{ 2\eta }{8\delta^3 } + 6\delta 2(\sqrt{2}+2\pi D)\D  f(\topt)
\end{align*}
Hence,
\begin{align*}
   f(\topt) -f(\tapp) & \le \frac{ 2\eta c}{8\delta^3 (f(\topt)+f(\tapp))} + \frac{6\delta 2(\sqrt{2}+2\pi D)\D  f(\topt)}{f(\topt)+f(\tapp)}  \\
   & \le \frac{ \eta c}{4\delta^3 f(\topt)} + \frac{12\delta (\sqrt{2}+2\pi D)\D  f(\topt)}{f(\topt)}  \qquad (\text{since } f(\tapp)>0) \\
   & \le \frac{ \eta c}{4\delta^3 \kappa\mA} + 12\delta (\sqrt{2}+2\pi D)\D   \qquad (\text{since } f(\topt)\ge \kappa\mA) \\
   & \le \frac{ \eta \mA\mB}{4\delta^3 \kappa} + 12\delta (\sqrt{2}+2\pi D)\D \qquad (\text{since } c\le \mA^2\mB \text{ by La.\ \ref{la:rm31}}) \enspace .
\end{align*}
Restricting this error to be at most $\e\mA$ and maximizing the expression for $\eta$ we get that the optimal values for $\delta$ and $\eta$ are
\begin{align}
   \label{eq:deta31}
   \delta&=\frac{\e  \mA}{16 (\sqrt{2}+2\pi D)\Delta}  \qquad \text{ and } \qquad    \eta  =\frac{\e^4 \kappa \mA^3}{16^3 \mB (\sqrt{2}+2\pi D)^3 \Delta^3} \enspace .
\end{align}
The probability of failure is $q=\frac{2}{3}M^3 e^{-\eta^2(M-3)/8}\le e^{-\eta^2 (M-3)/16}$. Analogous to the proof of Theorem~\ref{thm:main_absolute} we can show that $q\le \tau/2$ for all $M\ge \max\{-16/\eta^2 \log (2/\tau),\ 112/\eta^{2} \log (112/\eta^2) \}$.
Thus, if the number of recorded votes is \[M\ge k\cdot \frac{C'}{\e^8 \kappa^2}  \log \left( \max\left\{ \frac{1}{\tau},\ \frac{C'}{\e \kappa} \right\} \right)  \enspace ,\] where   $C'= \mB^2 \Delta^6 D^6 / \mA^{6} $,
$\Delta$ is the length of the boundary of $A$, $D$ is the diameter of $A$, and $k$ is an appropriate constant, then the area of overlap determined by the algorithm differes by at most $\e\mA$ from the maximal area of overlap with probability at least $1-\tau/2$.

Next we determine the total number $N$ of random samples that the algorithm needs to generate in order to record at least $M$ votes with probability at least $1-\tau/2$. For that purpose we first determine the probability that one randomly generated sample is not rejected. 

Let $\mathcal{C}_A$, and $\mathcal{C}_B$ denote the largest inscribed circles in  $A$ and $B$, respectively. And let $r_A, r_B$ denote the radii of $\mathcal{C}_A$ and $\mathcal{C}_B$. By the definition of $\kappa$-fatness $|\mathcal{C}_A|\ge \kappa\mA$ and $|\mathcal{C}_B|\ge \kappa \mB$, and by a precondition of the theorem $r_A\le r_B$.

Consider a circle $\mathcal{C}'_A$ of radius $r_A/2$ contained in $A$ and a circle $\mathcal{C}'_B$ concentric with $\mathcal{C}_B$ of radius $r_B-r_A/2\ge r_B/2$ in $B$. The area of $\mathcal{C}'_A$ is at least $\kappa\mA / 4$ and the area of $\mathcal{C}'_B$ is at least $\kappa\mB/4$. Then the probability that two randomly selected points $a_1$ and $a_2$ from $A$ are both contained in $\mathcal{C}'_A$ is at least $(\kappa/4)^2$. The distance between $a_1$ and $a_2$ is at most $r_A/2$. The probability that a randomly selected point $b_1$ from $B$ is contained in $\mathcal{C}'_B$ is at least $\kappa/4$, and by construction, the complete circle centered at $b_1$ with radius equal to the distance between $a_1$ and $a_2$ is completely contained in $\mathcal{C}_B\subset B$. Therefore, for every choice of two points in $\mathcal{C'}_A$ and a point in $\mathcal{C}'_B$ and for every randomly chosen direction the random sample induces a vote in transformation space. The probability that one random sample is not rejected is then at least as large as $(\kappa/4)^3$. In the following this probability is denoted by $p$.

Our algorithm generates in every step random samples independently. For each sample the probability not to be rejected is at least $p$. Then the expected number of valid samples after $N$ steps is at least $pN$. Let $X$ denote the number of valid samples after $N$ steps. Using the Chernoff bound (Theorem~\ref{thm:chernoff}) we can determine the number of steps $N$ for which $X$ is not much smaller than $pN$ with high probaility:
\begin{align*}
   P\left( X< pN - \xi N \right)  < 2e^{-\xi^2 N / 2}\enspace ,
\end{align*}
for all $0<\xi<1$. For $\xi=p/2$ we have $P\left( X< \frac{p}{2}N  \right)  < 2e^{-p^2 N / 8}$. Restricting this failure probability to be at most $\tau/2$ we get that for $N\ge 8p^{-2} \ln (4/\tau))$ the number of votes $X$ is at least $pN/2$ with probability at least $1-\tau/2$. Then with $N\ge 2M/p$ random samples the algorithm geenrates at least $M$ votes with probability at least $1-\tau/2$. 

Finally, choosing \[N\ge \max\left\{ \frac{2M}{p},\ \frac{8}{p^2} \ln (4/\tau)) \right\}= \max\left\{ k_1\cdot \frac{C'}{\e^8 \kappa^5}  \log \left( \max\left\{ \frac{1}{\tau},\ \frac{C'}{\e \kappa} \right\} \right), \ k_2\cdot\frac{1}{\kappa^6} \log \frac{1}{\tau} \right\}\]
with appropriate constants $k_1,k_2$, we get that with probability at least  $(1-\tau/2)$ the number of recorded votes ist sufficiently large, and therefore, with probaility at least $(1-\tau/2)^2\ge 1-\tau$ the approximation of the maximum area of overlap differs from the optimum by at most $\e\mA$.
\end{proof}

\comment{
\paragraph{proof thm 4}
}
\begin{proof}[Proof of Theorem \ref{thm:time}]
For a given arrangement $\mathcal{A}$ and an error tolerance $\e$ the depth approximation algorithm of \cite{approximateDepth} finds a point $\tdpt$ such that the depth of $\tdpt$  is at least $\e$ times the maximum depth in $\mathcal{A}$. It remains to show that  with high probability the area of overlap induced by $\tdpt$ is a good approximation of the maximal area of overlap. 

Let $h(t)$ be the density of the probability distribution $P$ in the transformation space induced by the algorithm, and $L$ the Lipschitz constant of $h$. Let $\mathcal{A}$ be the arrangement of $\delta$-neighborhoods of the transformations resulting from the random experiments for some parameter $\delta>0$ which will be specified later.  Additionally we will use an error tolerance value  $\eta>0$ when applying lemmas or corollaries. The parameter $\eta$ will also be computed later. 

      Let  $\tapp$ be a transformation with maximum depth in the arrangement $\mathcal{A}$, $\tdpt$ the transformation approximating the maximum depth, i.e., $X_N^{\delta}(\tdpt)\ge (1-\eta)X_N^{\delta}(\tapp)$, and $\topt$ the transformation maximizing $f(t)$. And let $q$ denote the failure probability from Lemma~\ref{la:key}.

Applying Lemma \ref{la:Lipschitz} and Corollary \ref{cor:output} to $\tdpt$ and error tolerance $\eta$ we get
\begin{align}
   |\md h(\tdpt)-X_N^{\delta}(\tdpt)/N| &\le |\md h(\tdpt)-P(\dBall[\tdpt])| + |P(\dBall[\tdpt]) - X_N^{\delta}(\tdpt)/N|  \notag\\
   &\le \eta+5L\md\delta 
   \label{eq:app}
\end{align}
and, likewise, by Lemmas \ref{la:Lipschitz} and \ref{la:key}
\begin{align}
   \vert \md h(\topt) - X_N^\d(\topt)/N \vert &    \leq  L \md \d +\eta \enspace . 
   \label{eq:opt}
\end{align}

Since $\mt[\tdpt] \leq \mt[\topt]$ holds by definition,
we only have to show $\mt[\tdpt] \geq \mt[\topt] - \e \mA$.

Consider Case 1: $X_N^{\delta}(\tdpt)\ge X_N^{\delta}(\topt)$. Then
\begin{align*}
   h(\tdpt) & \ge (X_N^{\delta}(\tdpt)-\eta -5L\md \delta) / \md \qquad \text{by \eqref{eq:app}}\\
   & \ge (X_N^{\delta}(\topt)-\eta -5L\md \delta) / \md \\
   & \ge (\md h(\topt)-\eta - L\md \delta -\eta -5L\md \delta) / \md \qquad \text{by \eqref{eq:opt}} \\
   & = h(\topt) - 2\eta / \md -6L \delta \enspace .
\end{align*}
If $X_N^{\delta}(\tdpt)< X_N^{\delta}(\topt)$ (Case 2)  we have that 
\[X_N^{\delta}(\tdpt)< X_N^{\delta}(\topt) \le X_N^{\delta}(\tapp) \quad \text{and}\quad |X_N^{\delta}(\tdpt)-X_N^{\delta}(\tapp)|\le \eta X_N^{\delta}(\tapp)\enspace . \]
It follows that also $|X_N^{\delta}(\tdpt)-X_N^{\delta}(\topt)|\le \eta X_N^{\delta}(\tapp)$.

Now we can bound the difference $|\md h(\tdpt)-\md h(\topt)|$ :
\begin{align*}
   |\md h(\tdpt)-\md h(\topt)| & \le |\md h(\tdpt)-P(\dBall[\tdpt])| + |P(\dBall[\tdpt])-X_N^{\delta}(\tdpt)/N|\\
   & \quad +|X_N^{\delta}(\tdpt)/N-X_N^{\delta}(\topt)/N|\\
   & \quad + |X_N^{\delta}(\topt)/N-P(\dBall[\topt])| + |P(\dBall[\topt])-\md h(\topt)| \enspace .
   \intertext{Applying Lemma \ref{la:Lipschitz} to the first term of the  expression, Corollary \ref{cor:output} to the second, Lemma \ref{la:key} to the fourth, Lemma \ref{la:Lipschitz} to the fifth, and bounding the third term by $\eta X_N^{\delta}(\tapp)/N\le \eta$ from above we get:}
   |\md h(\tdpt)-\md h(\topt)| &\le L\md\delta + \eta + 4L\md\delta + \eta + \eta + L\md\delta \\
   &= 3\eta + 6 L\md\delta \enspace.
\end{align*}
Then in Case 1 and in Case 2  $h(\tdpt)  \ge h(\topt) - 3\eta / \md -6L \delta $. Observe that this error bound for $\tdpt$ differs from the bound for $\tapp$ in Lemma~\ref{la:hApprox} only by a constant in the summand $\eta/\md$. Therefore we can complete the proof of the Theorem \ref{thm:time} by the same arguments as in the proof of Theorem~\ref{thm:main_absolute} with values of $\delta$ and $\eta$ that differ from Equations~\eqref{eq:dht}, \eqref{eq:dhrm}, and \eqref{eq:deta31} only by constants.
\end{proof}

\section{Open Questions}

We showed that our algorithm is a relative error approximation under the assumption that the maximal area of overlap of the two input shapes is at least a small constant fraction of one of the shapes. This is a reasonable assumption, but nevertheless it would be interesting to know whether the algorithm is a relative error approximation without further assumptions on the shapes.

Furthermore, it might be reasonable to use measure theoretic methods for the analysis of the algorithm. We can show that $X_{N}^\d(t)/N$ converges almost surely uniformly to $P(\dBall)$ on $T$, using the Uniform Law of Large Numbers II, as stated in \cite{hoffmann-jorgensen}. But it appears difficult  to determine the convergence rate to deduce bounds on the required number of random experiments by measure theoretic methods.

It is an interesting question how to apply our probabilistic technique to matching shapes under similarity maps. The straightforward technique of choosing a pair of random points from $A$ and one from $B$ and giving a vote to the corresponding similarity map does not lead to the desired result.

Furthermore, choosing three points in each shape defines a unique affine transformation that maps the points onto each other, so there is a canonical version of the algorithm for affine transformations. It would be interesting to know whether this algorithm computes something useful if the parametrization of the space and the definition of the neighborhoods are chosen cleverly.


\begin{thebibliography}{10}

\bibitem{Ahn:2008jx}
Hee-Kap Ahn, Siu-Wing Cheng, and Iris Reinbacher.
\newblock Maximum overlap of convex polytopes under translation.
\newblock In {\em 11th Japan-Korea Joint Workshop on Algorithms and
  Computation}, Fukuoka, 2008.

\bibitem{constant_factor}
Helmut Alt, Ulrich Fuchs, G{\"u}nter Rote, and Gerald Weber.
\newblock Matching convex shapes with respect to the symmetric difference.
\newblock {\em Algorithmica}, 21:89--103, 1998.

\bibitem{alt99discrete}
Helmut Alt and Leonidas~J. Guibas.
\newblock Discrete geometric shapes: Matching, interpolation, and
  approximation. a survey.
\newblock In J.-R. Sack and J.~Urrutia, editors, {\em Handbook of Computational
  Geometry}, pages 121--153, Amsterdam, 1999. Elsevier Science Publishers B.V.
  North-Holland.

\bibitem{09-smrs}
Helmut Alt and Ludmila Scharf.
\newblock Shape matching by random sampling.
\newblock In {\em 3rd Annual Workshop on Algorithms and Computation (WALCOM
  2009)}, volume 5431 of {\em Lecture Note in Computer Science}, pages
  381--393, 2009.

\bibitem{approximateDepth}
Boris Aronov and Sariel Har-Peled.
\newblock On approximating the depth and related problems.
\newblock In {\em SODA '05: Proceedings of the sixteenth annual ACM-SIAM
  symposium on Discrete algorithms}, pages 886--894. Society for Industrial and
  Applied Mathematics, 2005.

\bibitem{finding_a_guard}
Otfried Cheong, Alon Efrat, and Sariel Har-Peled.
\newblock Finding a guard that sees most and a shop that sells most.
\newblock {\em Discrete and Computational Geometry}, 37(4):545--563, 2007.

\bibitem{deBerg}
Mark de~Berg, Olivier Devillers, Marc~J. van Kreveld, Otfried Schwarzkopf, and
  Monique Teillaud.
\newblock Computing the maximum overlap of two convex polygons under
  translations.
\newblock {\em Theory of computing systems}, 31:613--628, 1998.

\bibitem{hoffmann-jorgensen}
J\o{}rgen Hoffmann-J\o{}rgensen.
\newblock {\em Probability with a view toward statistics, volume I \& II}.
\newblock Chapman \& Hall, New York, 1994.

\bibitem{krengel}
Ulrich Krengel.
\newblock {\em Einf\"uhrung in die Wahrscheinlichkeitstheorie und Stochastik}.
\newblock Vieweg, Braunschweig/Wiesbaden, fifth edition, 2000.

\bibitem{Veltkamp:fk}
Longin~Jan Latecki and Remco~C. Veltkamp.
\newblock Properties and performances of shape similarity measures.
\newblock In {\em Proceedings of International Conference on Data Science and
  Classification (IFCS)}, 2006.

\bibitem{mount}
David~M. Mount, Ruth Silverman, and Angela~Y. Wu.
\newblock On the area of overlap of translated polygons.
\newblock {\em Computer Vision and Image Understanding: CVIU}, 64(1):53--61,
  1996.

\end{thebibliography}
\end{document}